\documentclass[aps,prd,twocolumn,letterpaper]{revtex4}
\usepackage[dvips]{graphicx}
\usepackage{epsfig}

\def\p{\partial}
\def\d{{\rm d}}

\def\dd#1#2{\frac{\d #1}{\d #2}}
\def\pp#1#2{\frac{\p #1}{\p #2}} 
\def\rmmat#1{{\hbox{\rm #1}}}
\def\rmscr#1{\rmmat{\scriptsize #1}}
\def\acr#1{{\hat a}^\dag_{#1}}
\def\aan#1{{\hat a}_{#1}}

\begin{document}

\title{See a Black Hole on a Shoestring}
\author{Jeremy S. Heyl}
\address{Department of Physics and Astronomy, 
University of British Columbia \\ 
6224 Agricultural Road, Vancouver, British Columbia, Canada, V6T 1Z1;
Canada Research Chair}
\date{\today}

\begin{abstract}
  The modes of vibration of hanging and partially supported strings
  provide useful analogies to scalar fields travelling through
  spacetimes that admit conformally flat spatial sections.  This wide
  class of spacetimes includes static, spherically symmetric
  spacetimes.  The modes of a spacetime where the scale factor depends
  as a power-law on one of the coordinates provide a useful starting
  point and yield a new classification of these spacetimes on the
  basis of the shape of the string analogue. The family of
  corresponding strings follow a family of curves related to the
  cycloid, denoted here as hypercycloids (for reasons that will become
  apparent).  Like the spacetimes that they emulate these strings
  exhibit horizons, typically at their bottommost points where the
  string tension vanishes; therefore, hanging strings may provide a
  new avenue for the exploration of the quantum mechanics of horizons.
\end{abstract} 

\maketitle

\section{Introduction}

The rules governing the interaction of black holes in relativity bears a
remarkable similarity to the laws of thermodynamics established with
laboratory experiments \cite{Bekenstein:1973ur,Hawking:1974sw}.  Over
the past several years there has been great excitement about bringing
the resemblance full circle with the description and realization of
laboratory analogues to the structure of curved spacetimes
\cite{1981PhRvL..46.1351U,volovik03:_univer_helium_dropl}. Studying
these analogues has given new insight on how the properties of black
hole horizons probe the underlying theory of quantum gravity
\cite{2005PhRvD..71b4028U}.  These laboratory systems have generally
involved surface or pressure oscillations of superfluids
\cite{volovik03:_univer_helium_dropl} and Bose-Einstein condensates
\cite{2003PhRvA..68e3613B}, and more recently microwaves in specially
designed waveguides \cite{2005PhRvL..95c1301S}.

Although there has been substantial work on the dynamics of acoustic
oscillations near vortexes, much of the effort to understand
acoustical analogues to spacetimes has focused on $(1+1)$-dimensional
systems \cite[e.g.]{barcelo04:_causal} that are either static or
stationary.  By dimensional reduction (focusing only on the radial
modes), a general spherically symmetric spacetime in
$(3+1)$-dimensional can be represented by a $(1+1)$-dimensional
analogue \cite{2005CQGra..22..409C}.  This letter examines possibly
the simplest realization of a $(1+1)$-dimensional analogue to
spacetime, transverse oscillations travelling along a string --- this
is a layperson's string like a guitar or violin string rather than a
fundamental or cosmic string.  By varying the tension along the
string, one can tune the properties of the analogue spacetime and even
produce acoustic horizons along the string.




\section{Waves on a String}

The action for oscillations on a string with varying tension and mass
density is given by \cite[][p. 597, 598]{Gold80,Gold02}
\begin{eqnarray}
S &=& 
\frac{1}{2} \int ds dt \left [ 
\mu(s) \left ( \pp{\phi}{t} \right )^2 - \lambda(s) \left ( \pp{\phi}{s}
\right )^2 \right ] \label{eq:5} \\ 
&=&
\frac{1}{2} \int ds dt \mu(s) \left [ 
 \left ( \pp{\phi}{t} \right )^2 - c_s^2(s)\left ( \pp{\phi}{s}
\right )^2 \right ]  
\label{eq:6}
\end{eqnarray}
where $\mu$ is the mass per unit length of the string,
$c_s^2=\lambda/\mu$ and $\phi$ is the displacement of the string in
one direction, and the tension $\lambda(s)$ is not affected by the
oscillations.  Let's define $c_s d \chi = ds$ and change variables in
the integral to get
\begin{equation}
S =
\frac{1}{2} \int d\chi dt a^2(\chi) \left [ 
 \left ( \pp{\phi}{t} \right )^2 - \left ( \pp{\phi}{\chi}
\right )^2 \right ]  
\label{eq:7}
\end{equation}
where $a^2(s)=c_s(s)\mu(s)$.

This action for the acoustic modes along a string is the same as the
action for a scalar field in a conformally flat spacetime.
\S~\ref{sec:string-horizons} will extend this result to spacetimes
with conformally flat {\bf spatial} sections.

Let's make the substitution $u=a\phi$ to try to absorb the
factor of $a^2(\chi)$ into the fields.
\begin{eqnarray}
S &=& 
\frac{1}{2} \int d \chi d t  \left [ 
\left ( \pp{u}{t} \right )^2 - 
\left ( a \pp{(u/a(\chi))}{\chi} \right )^2
\right ]
\label{eq:8}
\\ 
&=& \frac{1}{2} \int d\chi d t  \Biggr [ 
\left ( \pp{u}{t} \right )^2 - \left (\pp{u}{\chi} \right)^2   -
m_\rmscr{eff}^2 u^2
\Biggr ] 
\label{eq:9}
\end{eqnarray}
where to get the final result we have dropped a term in the integrand 
equal to a total derivative with respect to the coordinate $\chi$,
specifically we take
\begin{equation}
\frac{1}{2} \int d\chi d t \dd{}{\chi} \left ( \frac{u^2}{a}
\pp{a}{\chi} \right ) = 0
\label{eq:10}
\end{equation} 
We find that the oscillations of the string acquire an effective mass of
\begin{equation}
m^2_\rmscr{eff}=\frac{1}{a} \pp{^2 a}{\chi^2}.
\label{eq:11}.
\end{equation}

\subsection{Quantizing the String}
\label{sec:quantizing-string}

The first step to determining the quantum mechanics of the string is to
obtain the Hamiltonian,
\begin{equation}
H = \frac{1}{2} \int d\chi  \left [ 
\left ( \pp{u}{t} \right )^2 + \left (\pp{u}{\chi} \right)^2   + m_\rmscr{eff}^2  u^2
\right ] 
\label{eq:12}
\end{equation}
To look at the quantum mechanics of this Hamiltonian we must define
the field operator ${\hat u}(\chi,t)$,
\begin{equation}
{\hat u}(\chi,t) = \int d \omega \left [ g(\omega,\chi) e^{-i\omega t} \aan{\omega} + 
g^*(\omega,\chi) e^{i\omega t} \acr{\omega} \right ]
\label{eq:13}
\end{equation}
where $\acr{\omega}$ and $\aan{\omega}$ are creation and annihilation
operators that satisfy the following standard commutator relations
\begin{equation}
\big [ \acr{\omega}, \acr{\omega'} \big ] = 
        \big [ \aan{\omega}, \aan{\omega'} \big ] = 0,
        \big [ \aan{\omega}, \acr{\omega'} \big ] = 
                        \delta ({\omega} - {\omega'}).
\label{eq:14}
\end{equation}
Because the Hamiltonian does not explicitly depend on time, we are
free to choose that the field operator depend harmonically on time.

If we assume that the function $g(\omega,\chi)$ solves the following
differential equation
\begin{equation}
 \pp{^2 g}{\chi^2} + \left (\omega^2 - m^2_\rmscr{eff}  \right ) g = 0
\label{eq:15}
\end{equation}
then the Hamiltonian operator takes the simple form
\begin{equation}
{\hat H} = \frac{1}{2} \int d \chi d \omega \omega \left (
\aan{\omega}\acr{\omega} +
\acr{\omega} \aan{\omega} \right )
\label{eq:16}
\end{equation}

\subsection{Linear Tension}

Let's calculate $a(s)=\sqrt{c_s(s) \mu(s)}$ if the string is hanging
vertically with one end of the string loose at $s=0$ and $\mu(s) = b
s^\alpha$.  We have 
\begin{equation}
\lambda(s)= \int_0^s g \mu(s') ds' = \int_0^s g b \left (s'\right )^\alpha d s' = \frac{b s^{\alpha+1}}{\alpha+1},
\end{equation}
so
\begin{equation}
c_s = \sqrt{ \frac{g s}{\alpha+1}}, \chi = 2
\sqrt{\frac{s}{g}(\alpha+1)}.
\label{eq:17}
\end{equation}
for $\alpha>-1$.  Putting things together we have
\begin{equation}
a(\chi) = b^{1/2} \left ( \frac{\chi}{2} \right )^{\alpha+1/2} 
\left ( \frac{g}{\alpha+1} \right )^{(\alpha+1)/2} 
\label{eq:18}
\end{equation}
and 
\begin{equation}
m^2_\rmscr{eff}=\frac{1}{\chi^2} \left ( \alpha^2- \frac{1}{4} \right )
\label{eq:19}\end{equation}
The coordinate $\chi$ measures the time that it takes a wave to reach
a distance $s$ along the string from the free end.  Even though the
sound speed vanishes at $s=0$, the time for sound waves to come from
the free end is finite.  Substituting this result into
Eq.~(\ref{eq:15}) yields
\begin{equation}
 \pp{^2 g}{\chi^2} + \left (\omega^2 - \frac{1}{\chi^2} \left ( \alpha^2- \frac{1}{4} \right ) \right ) g = 0
\label{eq:20}
\end{equation}
Substituting $g(\chi)=\sqrt{\chi}h(\chi)$ gives Bessel's equation
\begin{equation}
\chi^2  \pp{^2 h}{\chi^2} + \chi \pp{h}{\chi} + \left (\chi^2 \omega^2 - \alpha^2  \right ) h = 0
\label{eq:21}
\end{equation}
so we have
\begin{equation}
g(\chi) = \sqrt{\chi} \left [ c_1 J_\alpha (\omega \chi) + c_2
Y_\alpha(\omega \chi) \right ]
\label{eq:22}
\end{equation}
Because the string has a finite length $L$, the variable $\chi$ varies
from $\chi_1=0$ to 
\begin{equation}
\chi_2 = 2  \sqrt{\frac{L}{g}(\alpha+1)}
\label{eq:23}
\end{equation}
How the string is attached at $L$ determines a boundary condition 
at $\chi_2$.  For example, if the string is fixed at $L$, $\chi_2$
must be a zero of the function $g(\chi)$.  At the free end
($\chi_1=0$) the displacement of the string must not diverge so
$c_2=0$.   These two boundary conditions determine the spectrum of 
values for $\omega$.   

\subsection{Power-Law Tension}

Let's assume a more complicated dependence of the force on the
distance along the string.  Specifically let us take
\begin{equation}
d \lambda =  q s^\beta \mu d s = q b s^{\alpha+\beta} d s
\label{eq:24}
\end{equation}
so
\begin{equation}
\lambda(s) = \frac{q b}{\alpha+\beta+1} s^{\alpha+\beta+1}, 
c_s = \sqrt{\frac{q s^{\beta+1}}{\alpha+\beta+1}}
\label{eq:25}
\end{equation}
The coordinate $\chi$ is now a bit more complicated
\begin{equation}
\chi = \frac{2}{1-\beta} \sqrt { \frac{\alpha+\beta+1}{q } } 
s^{(1-\beta)/2} + C
\label{eq:26}
\end{equation}
If we take $\beta=0$ and $q=g$, we obtain Eq.~(\ref{eq:17}), 
and if $\beta=1$ we have
\begin{equation}
\chi = \sqrt{\frac{2+\alpha}{q}} \ln s + C
\label{eq:27}
\end{equation}
Specifically if $\beta \geq 1$ it takes a infinite amount of time for
a wave to propagate from the free end of the string upward.  The free
end of the string is a horizon for waves travelling along the string.
For $\beta < 1$ is the natural to set the arbitrary constant of
integration $C$ equal to zero, so that $\chi$ gives the amount of time
for a excitation at the end of the string to reach a point a distance
$s$ away.  For $\beta \geq 1$, $C$ may be defined conveniently by
taking the zero point of $\chi$ to be the fixed end of the string $s=L$.

In general we have
\begin{equation}
a(s) = \left ( \frac{b^2 q}{\alpha+\beta+1} \right )^{1/4} s^{(2\alpha+\beta+1)/4}.\label{eq:28}
\end{equation}
and
\begin{equation}
a(\chi) = \left ( \frac{b^2 q}{\alpha+\beta+1} \right )^{1/4}
\left ( \frac{1-\beta}{2} \sqrt{\frac{q}{\alpha+\beta+1}} \chi \right
)^{\frac{2\alpha+\beta+1}{2-2\beta}}
\label{eq:29}
\end{equation}

For $\beta\neq 1$ we have
\begin{equation}
m^2_\rmscr{eff} = \frac{1}{\chi^2}
\frac{(2\alpha+3\beta-1)(2\alpha+\beta+1)}{4(1-\beta)^2}.
\label{eq:33}
\end{equation}
The effective mass-squared is bounded from below by $-1/(4\chi^2)$ just as for 
Eq.~(\ref{eq:11}) so the solutions developed earlier apply for
arbitrary values of $\beta$ but with different boundary conditions if
$\beta \geq 1$.  If $\beta \geq 1$, the end of the string ($s=0$)
corresponds to $\chi \rightarrow -\infty$, so one gets a continuous
spectrum of eigenvalues $\omega$.

For $\beta=1$ we have
\begin{equation}
m^2_\rmscr{eff} = q \frac{\alpha+1}{4}
\label{eq:34}
\end{equation}
In this special case, the function $g(\omega,\chi)$ does not satisfy
Bessel's equation but a simpler version of Eq.~(\ref{eq:15}) with a
constant value of $m^2_\rmscr{eff}$.  Here both the energy and the 
$\chi-$momentum of the modes are conserved and we have
\begin{equation}
{\hat u}(\chi,t) =  e^{-i(k\chi - \omega t)} \aan{\omega} + 
e^{i(k\chi - \omega t)} \acr{\omega}
\label{eq:35}
\end{equation}
with $\omega^2=k^2+m^2_\rmscr{eff}$.  Although the picture looks
simple in terms of the $\chi$ variable, the reality is a bit more
complicated.  The string has a finite length over which $\chi$ runs
from $-\infty$ to 0 from Eq.~(\ref{eq:27}).

\subsection{Setting the Tension}

Fixing the tension in the string to follow an arbitrary function is
actually less complicated than it might first appear.  Because the
analysis is concerned with oscillations in only a single direction (to
be definite call this the $y-$direction), the string can be forced to
lie upon a particular surface $z=z(x)$; thereby, setting the tension.
If we resolve the downward acceleration of gravity ($g$) along the
direction of the string, we find that the tension increases along the
string as
\begin{equation}
d \lambda = \mu \cos \theta g ds = \mu g \frac{dz}{ds} ds = \mu g dz
\label{eq:36}
\end{equation}
where $\theta$ is the angle that the string makes with the vertical at
a particular location.  If the mass-density along the string is
constant ($\alpha=0$), we have
\begin{equation}
\lambda = \mu g z = \frac{q \mu}{\beta + 1} s^{\beta+1}
\label{eq:37}
\end{equation}
where have used Eq.~(\ref{eq:25}).  Two immediate solutions come to
mind.  First, the trivial solution for $\beta=0$ is simply $z=s$, the
string hangs vertically.   Second, $\beta=1$ yields
\begin{equation}
g z = \frac{q}{2} s^2.
\label{eq:38}
\end{equation}
If one looks at a particle sliding along a curve that satisfies
Eq.~(\ref{eq:38}), the potential energy is proportional to the square
of the displacement -- so the frequency of the oscillatory motion is
independent of the amplitude.  The curve with this property is called
the tautochrone or isochronous.  Huygens \cite{Huyg73} found that the
cycloid satisfies Eq.~(\ref{eq:38}) (also see \cite{Melv99}),
\begin{equation}
  x = \frac{g}{8q} \left ( \theta - \sin \theta \right ), z =
  \frac{g}{8q} (\cos \theta - 1).
\label{eq:39}
\end{equation}
For general values of $\alpha$ and $\beta$ the curve must satisfy the
following differential equations
\begin{equation}
\dd{z}{s} = \frac{q}{g} s^\beta, \dd{x}{s} = \sqrt{1 - \frac{q^2}{g^2} s^{2\beta}}.\label{eq:40}
\end{equation}
The first equation results from combining Eq.~(\ref{eq:24}) and
Eq.~(\ref{eq:37}) -- the dependence on the mass density
vanishes.  The second equation results from the Pythagorean theorem.

The solution to these equations is 
\begin{eqnarray}
z &=& \frac{s}{(\beta+1)} \left (  \frac{q}{g} s^\beta \right ), 
\label{eq:41}
\\
x &=& s\,
_2F_1 \left ( \left [ -\frac{1}{2},\frac{1}{2\beta} \right ], \left [
  1+\frac{1}{2\beta} \right ] , 
\left ( \frac{q}{g} s^\beta \right )^2 \right )
\label{eq:42}
\end{eqnarray}
where $_pF_q$ denotes the generalized hypergeometric function.  Let's
denote this family of parametric curves, hypercycloids.  The value $\beta=1$
yields the usual cycloid and the value $\beta=0$, a vertical line.

\begin{figure}
\centerline{\includegraphics[height=2.2in]{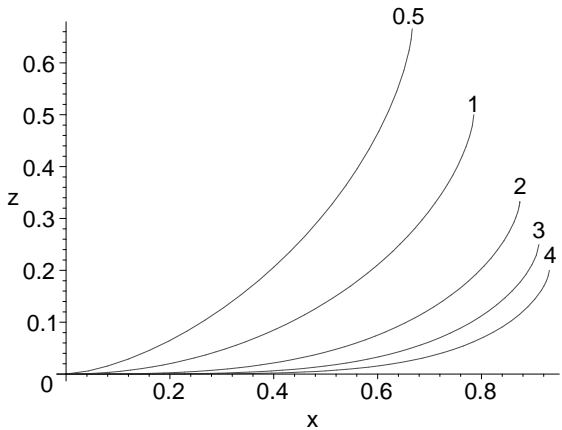} }
\smallskip 
\caption{The cross-section of the confining surfaces, hypercycloids
 for $\beta=0.5,1$ (the cycloid), $2,3$ and 4 from left to right. The curves are
 labelled with the value of $\beta$}
\label{fig:curves}
\end{figure}

Fig.~\ref{fig:curves} depicts the hypercycloids from $\beta=0.5,1,2,3$ and
4.  The curves are symmetric about the $z=0$ line and can be
scaled in the $x-$ and $z-$direction simultaneously.  This changes the
value of $g/q$.  However, the two directions cannot be scaled
independently. The cycloid ($\beta=1$) is the critical curve that
admits a horizon to string vibrations at $x=z=0$.

If the tension is some more general function $\lambda(s)$, we can use
Eq.~(\ref{eq:36}) to determine the necessary curve through the
following equations
\begin{equation}
\dd{z}{s} = \frac{1}{\mu g} \dd{\lambda}{s}, \dd{x}{s} = 
\sqrt{1-\left (\dd{z}{s}\right)^2}
\label{eq:43}
\end{equation}
Because $a^2(s)=c_s(s)\mu(s)=\sqrt{\lambda(s)\mu(s)}$,
Eq.~(\ref{eq:43}) allows us to build a string analogue to a wide
variety of conformally flat spacetimes.

\section{String Horizons}
\label{sec:string-horizons}

In the previous section, we characterize the mechanics of
oscillations of a string with a varying mass density and tension in terms of 
an effective acoustic metric. It is possible to
identify the conformally flat spacetime that describes the motion of
waves along a string with spherically symmetric scalar fields in
static, spherically symmetric spacetimes such as black holes.

The spacetime of a static, spherically symmetric system may be
described using isotropic coordinates \cite{Misn73}
\begin{equation}
ds^2 = B^2({\bar r}) dt^2 - A^2({\bar r})\, d {\bf \bar x}^2 
\label{eq:44}
\end{equation}
where ${\bar r}^2={\bar x}^2+{\bar y}^2+{\bar z}^2$.  The spatial
portion of this metric is conformally flat.

Let's write the action for a scalar field in this spacetime,
\begin{eqnarray}
S &=& 
\frac{1}{2} \int d^4 x \sqrt{-g} \left [ \partial_\mu  \phi
  \partial^\mu \phi \right ] \label{eq:45} \\
&=& 
\frac{1}{2} \int d^4 x A^3({\bar r}) B({\bar r})  \left [ 
\frac{1}{B^2({\bar r})} \left ( \pp{\phi}{t} \right )^2 - \frac{ \left (\nabla  \phi
\right)^2}{A^2({\bar r})} \right ] ~~~
\label{eq:46}
\end{eqnarray}
where we have taken $\hbar=c=1$.

Let's focus on spherically symmetric waves to get
\begin{eqnarray}
S &=& \frac{1}{2} \int dt {\bar r}^2 d{\bar r} d\Omega  \Biggr [ 
\frac{A^3({\bar r})}{B(\bar r)} \left ( \pp{\phi}{t} \right )^2
\nonumber \\ 
& & ~~~~~~~ - 
A(\bar r) B(\bar r)  \left
(\pp{\phi}{\bar r}
\right)^2 \Biggr ]  \label{eq:47} 
\end{eqnarray}
The dynamics of a spherically symmetric scalar field in this spacetime
can be mimicked by making the identifications
\begin{equation}
s={\bar r}, \mu(s) = r^2 \frac{A^3(\bar r)}{B(\bar r)}, 
\lambda(s) = {\bar r}^2 A(\bar r) B(\bar r).
\label{eq:48}
\end{equation}
Using Eq.~(\ref{eq:43}) we have 
\begin{equation}
\dd{z}{s}=\frac{1}{\mu g} \dd{\lambda}{s} = \frac{1}{r g A^2(\bar r) }
\left \{ 2 + \dd{\ln \left [ A(\bar r)B(\bar r) \right ]}{\ln r}
\right \}
\label{eq:49}
\end{equation}
with units with $G=c=1$.  Hence the product $rg$ is dimensionless, as
are the functions $A(\bar r)$ and $B(\bar r)$.
Although this equation may be solved for any spacetime
(Eq.~(\ref{eq:44})), it is useful to look at the form of this equation
in the vicinity of a horizon and compare it with Eq.~(\ref{eq:40}).

For example let's take the spacetime of a charged, non-rotating black
hole (the Reissner-Nordstr\"om metric)
\cite{reissner16:_ueber_eigen_feldes_einst_theor,nordstroem18:_einst,weyl17:_zur_gravity}, we have \cite{Misn57,1972PhRvD...5.1897B}
\begin{eqnarray}
A(\bar r) &=& \left ( 1 + \frac{M}{2{\bar r}}\right )^2 -
\frac{Q^2}{4{\bar r}^2} \label{eq:50} \\
B(\bar r) &=& \frac{1}{A(\bar r)} \left [ 1 + \frac{1}{4r^2} \left (
  Q^2 - M^2\right ) \right ]\label{eq:51}
\end{eqnarray}
where we have taken $G=c=1$.

In this case we have
\begin{equation}
\dd{z}{s}=\frac {32 r^5}{g \left( {\bar r}^{2}- {\bar r}_H^2 \right)
 \left[ \left ( 2{\bar r}+ M \right )^2 + Q^{2} \right] ^{2}}. \label{eq:52}
\end{equation}
This expression is actually integrable in terms of elementary
functions.  However, the corresponding $\d x/\d s$ is not.  The
horizon is located where $B(r)$ vanishes, that is at
\begin{equation}
{\bar r}^2 = {\bar r}_H^2 = \frac{M^2-Q^2}{4}. \label{eq:53}
\end{equation}
In general we find that $\d z/\d s$ diverges  as $({\bar r} - {\bar
  r}_H)^{-1}$
when ${\bar r} \rightarrow
{\bar r}_H$; this is difficult to
achieve as $|\d z/\d s| \leq 1$ by the Pythagorean theorem.  However,
if we look at an extremal Reisser-Nordstr\"om metric ($M^2=Q^2$), we have
$A(r)=1+M/{\bar r}$ and $B(r)=1/A(r)$ from Eq.~(\ref{eq:50}) and Eq.~(\ref{eq:51}) so
\begin{equation}
\lambda (s) = {\bar r}^2, \mu(s) = {\bar r}^2 \left ( 1 + \frac{M}{\bar r} \right )^4
\label{eq:54}
\end{equation}
and 
\begin{equation}
\dd{z}{s} = \frac{2 {\bar r}^3}{g \left ( {\bar r} + M \right )^4}
\label{eq:55}
\end{equation}
The expression $\d z/\d s$ reaches a maximum at $r=3M$ of $27/(2gM)$,
so if $M>27/(2g)$ we can construct a curve to represent the entire
spacetime; otherwise we are restricted to the region with $\d z/\d s
\leq 1$.

In practice however, the value of the gravitational acceleration on
Earth is so small ($\sim 10^{-16}$m$^{-1}$) that $M$ must be greater
than $10^{17}$m to emulate the entire extremal Reissner-Nordstr\"om
spacetime, so terrestrial experiments will necessarily focus on 
$r \ll\ M$, the region near the horizon and we have
\begin{equation}
\dd{z}{s} \approx \frac{2 {\bar r}^3}{gM^4}.
\end{equation}
This yields the hypercycloid with $\beta=3$ 
Unfortunately the mass density diverges when $r\rightarrow 0$ as
\begin{equation}
\mu(s) \approx \frac{M^4}{{\bar r}^2}
\end{equation}
so emulating even the region of this spacetime near the horizon may be
difficult.  Perhaps other spacetimes that lie at the critical point of
having a horizon may also exhibit simple string analogues like
Reissner-Nordstr\"om.

\section{Prospects}

The theoretical framework outlined here is not complete.
Specifically, the Lagrangian, Eq.~(\ref{eq:5}), assumes that the
presence of the oscillation does not affect tension in the string.
Although recent work \cite{2005PhRvD..71b4028U} indicates that the
interesting properties of horizons such as horizon radiation
\cite{Hawking:1974sw} are robust with respect to general assumptions
of the dispersion relation at high frequencies, the change in the
tension in the string due to the oscillation itself is a form of
non-linear back-reaction that is not covered by this analysis.  The
string may offer an avenue to explore both quantum and classical
back-reactions on otherwise static spacetimes.  Additionally the
dispersion relation for waves on a string is probably more complicated
that that described by Eq.~(\ref{eq:5}), so the issue discussed in
ref. \cite{2005PhRvD..71b4028U} may also be important.

The dynamics of strings has often presented surprises.  A common
example is the cracking of a bullwhip \cite{2002PhRvL..88x4301G} that
indicates that the speed of the tip of the whip has exceeded that of
sound. This letter focuses on the classical and quantum dynamics of
freely hanging and supported tapered strings (the bullwhip is a
particular case).  The acoustics of a waves along the string naturally
admits a Lorentzian description by means of an acoustic metric.  By
varying the tension or the mass-density of the string, the acoustics
can emulate the causal structure of variety of spacetimes.  This
letter has focused on conformally flat spacetimes
and spherically symmetric black holes in $(3+1)$ dimensions.
It is straightforward theoretically to produce acoustic horizons on
strings (a string resting on a concave-up cycloid is a limiting
case); perhaps, the experimental realization of such systems will
provide new insights into the small-scale structure of spacetime.

\acknowledgements

The author acknowledges support from NSERC. This work made use of
NASA's Astrophysics Data System.

\bigskip

\bibliographystyle{prsty}
\bibliography{mine,gr,inflation,qed,physics}

\end{document}